\documentclass{article}[15pt]

\usepackage{amsmath,epsfig,color,calc,rotating}
\usepackage{graphics}
\usepackage{wrapfig}
\usepackage{ulem}

\usepackage{natbib}
\bibpunct{(}{)}{;}{a}{}{,}

\begin{document}

\title{
Effective Sample Size: Quick Estimation of the Effect of 
Related Samples in Genetic Case-Control Association Analyses 
\author{
Yaning Yang$^{1}$, Elaine F. Remmers$^{2}$, Chukwuma B. Ogunwole$^{2}$, \\
Daniel L. Kastner$^{2}$, Peter K. Gregersen$^{3}$, Wentian Li$^{3}$ \\
{\small \sl 1. Department of Statistics and Finance,
University of Science and Technology of China, Anhui 230026, Hefei, CHINA} \\
{\small \sl 2. Genetics and Genomic Branch, National Institute of Arthritis and Musculoskeletal and Skin Diseases}\\
{\small \sl National Institute of Health, 9 Memorial Drive, Bethesda, MD 20892, USA.} \\
{\small \sl 3. The Robert S. Boas Center for Genomics and Human Genetics,  The Feinstein Institute }\\
{\small \sl for Medical Research, North Shore LIJ Health System, Manhasset, 350 Community Drive, NY 11030, USA.}\\
}
\date{}
}
\maketitle  
\markboth{\sl Yang, Gregersen, Li) }{\sl Yang, Gregersen, Li}

\begin{center}
{\bf Summary }
\end{center}

Affected relatives are essential for pedigree linkage analysis,
however, they cause a violation of the independent sample assumption
in case-control association studies.  To avoid the correlation
between samples, a common practice is to take only one affected
sample per pedigree in association analysis. Although several methods 
exist in handling correlated samples, they are still not widely used 
in part because these are not easily implemented, or because they
are not widely known. We advocate the effective sample size method as 
a simple and accessible approach for case-control association analysis with
correlated samples. This method modifies the chi-square test 
statistic, $p$-value, and 95\% confidence interval of the odds-ratio 
by replacing the apparent number of allele or genotype 
counts with the effective ones in the standard formula, without
the need for specialized computer programs. We present a  simple
formula for calculating effective sample size for many types of relative
pairs and relative sets. For allele frequency estimation, the 
effective sample size method captures the variance inflation 
exactly. For genotype frequency, simulations showed that effective 
sample size provides a satisfactory approximation.  A gene which is 
previously identified as a type 1 diabetes susceptibility locus, the 
interferon-induced helicase gene ({\sl IFIH1}), is shown to be significantly 
associated with rheumatoid arthritis when the effective sample size
method is applied. This significant association is not established 
if only one affected sib per pedigree were used in the association
analysis.   Relationship between the effective sample size
method and other methods -- the generalized estimation equation, 
variance of eigenvalues for correlation matrices, and genomic controls
--  are discussed.

\large

\section*{Introduction}

\indent

One of the major obstacles in statistical analysis of genetic
association studies in a case-control setting
\citep{lewis,balding,li-bib} is the violation
of the independence assumption. 
Dependence between samples, such as members from the same
family, invalidates a basic assumption in many statistical tests,
thus potentially making  the $p$-value estimation unreliable. 

As dependence has been an important theme in statistics
for many years, there is large amount of literature in genetics as well
as in statistics to tackle the problem. For example, the maximum 
likelihood or Bayes estimation of allele frequencies in relatives
\citep{boehnke,thomas,coram}; the use
of principal components or eigenvectors to identify clusters of 
samples \citep{price,patterson}, or the
reduction of  effective number of markers in a linkage disequilibrium block 
\citep{cheverud,dale}; sample weighting to suppress contributions 
from correlated samples \citep{broman,browning}, etc. 

The transition from genetic linkage analyses to association studies
\citep{risch,lim} presents a situation
when affected sibs or affected pedigree members are often used as case
samples in a case-control association study
\citep{bourgain05,epstein,moore,biedermann,klei,kohler07,yoo,visscher,camp09}. 
Since the correlation
structure between sibs or relatives is given, it is not necessary to
use techniques such as the generalized estimating equation as has been
carried out in \citep{silverberg}.  Instead, variance of correlated samples
can be calculated \citep{slager} and its effect on the test
statistic can be determined. The method discussed in
\citep{slager} is however only applied to the Armitage trend test.

\begin{figure}[t]
\begin{center}
  \begin{turn}{-90}
  \epsfig{file=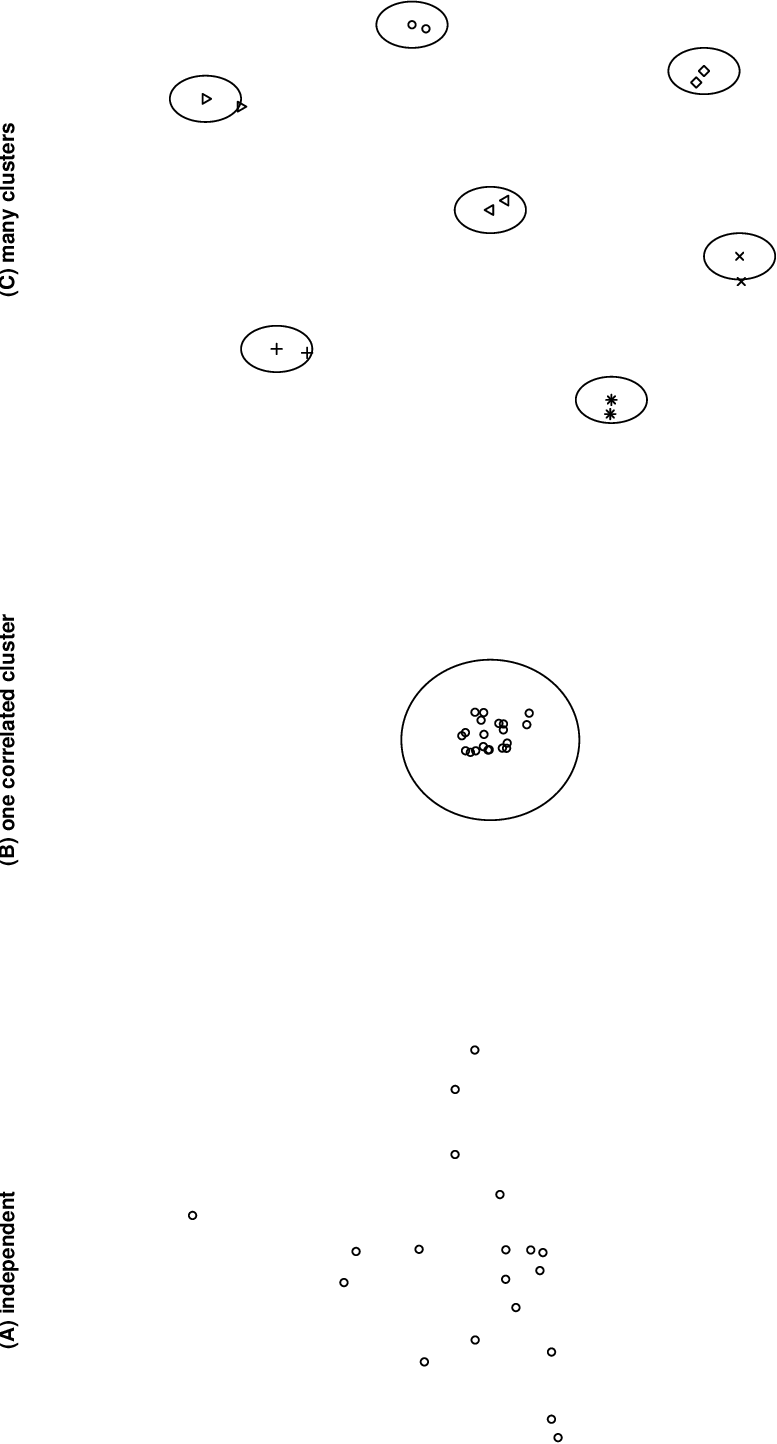, width=6.5cm}
  \end{turn}
\end{center}
\caption{
\label{fig:illus}
Illustration of three situations concerning sample correlations:
(A) samples are independent;
(B) all samples are correlated with each other to form one cluster;
(C) samples within a cluster are correlated, whereas there is no
correlation between clusters. This is called ``cluster-correlated
data" in \citep{williams}.
}
\end{figure}

To avoid confusion, Fig.\ref{fig:illus} illustrates the
situation to be addressed in this paper.  Fig.\ref{fig:illus}(A) is the
standard situation where samples are independent. Fig.\ref{fig:illus}(B)
shows the situation where all samples are correlated with one another.
This is however not the situation we will address. Fig.\ref{fig:illus}(C) consists
of correlated clusters, whereas there is no correlation between clusters
themselves. Fig.\ref{fig:illus}(C) is the situation when relatives
of the same family are used for association analysis.

Fig.\ref{fig:illus}(B) leads to a smaller variance
compared to independent situation Fig.\ref{fig:illus}(A) with the
same number of samples. Since larger sample size leads to smaller
variance, it is as if the ``effective sample size" is increased
in Fig.\ref{fig:illus}(B). The trend in Fig.\ref{fig:illus}(C) is the opposite:
the ``effective sample size" is actually reduced.
Take an extreme example of monozygotic twins: since monozygotic twins have identical
genotypes, a pair of twins provide the same genetic information as one twin,
and the two points within a circle in Fig.\ref{fig:illus}(C) is
equivalent to one point. In other words, the effective sample
size is only half of the apparent sample size.
These concepts have already been understood in the study of clustered/clumped data
and are associated with phrases like ``variance inflation"
and ``overdispersion".

In this paper, we advocate the use of ``effective sample size" (ESS) as a simple
method to capture the effect of sample correlation and variance inflation.
The term effective sample size has appeared in the literature before
\citep{kish,thiebaux,rosner,rao,madden} but has not become a commonly
used tool in genetic analysis. We define effective sample size $N_E$
as the equivalent number of independent samples that leads to the
same variance of an {\sl intensive quantity}, i.e., a quantity that does not
change with the sample size. 

For example, if the
sample proportion of heads in a coin tossing is estimated to be $p$,
its variance is $p(1-p)/N$ where $N$ is the number of coin tosses;
if the observed variance is larger than what is expected from
this equation, and can be fitted by the formula $p(1-p)/N_E$, then
$N_E$ is the effective sample size. Note that this definition of $N_E$
is very similar to the ``variance effective size" used in population
genetics, but different from, and should not be confused with,
the ``inbreeding effective population size" ($N_e$) also used 
in population genetics \citep{wright}.

In genetic case-control studies, the association signal originates
from the allele or genotype frequency difference in the diseased and
the normal group. The estimation of allele or genotype frequency is very
much like the estimation of heads proportion in the tossing coin example
given above.  We will show that for allele frequency, effective sample size
captures the effect of sample correlation exactly. Even for situations
where the effective sample size does not provide an exact solution,
for example, in estimating genotype frequencies, an averaged parameter
usually leads to good approximation.  Because the calculation
of test statistics $X^2$, $p$-value, and power all directly involve sample
size, replacing the apparent sample size with the effective sample
size is a quick and convenient solution to the problem of correlated samples
without the need to use a custom program.

As there are many publications on the effect of sample correlation on 
association analysis, and on using pedigrees in association studies, 
related questions that are not addressed here include:
(1) combining linkage and association signals \citep{joe4,lim};
(2) family-based associations such as transmission disequilibrium test 
(TDT) and its extensions \citep{nagel,camp05,gray};
(3) association using multiple family members with novel
test statistics instead of the standard chi-square test
\citep{teng1,teng2,zli};
(4) association with unknown (``cryptic") correlations
\citep{voight,astle,rakovski,mcpeek10,sillanpaa} where the relativeness between samples is
detected instead of given \citep{weir06,weir09}.
This paper is about a simple and accessible method to incorporate
sample correlations in genetic case-control studies 
within the standard chi-square test framework.

\section*{Mathematical Details}

\subsection*{Effective sample size for sibpairs} 

For simplicity, let's first consider $N_{\rm sib}$ sibpairs.
For the quantity of interest $x_i$ ($i=1,2, \cdots 2N_{\rm sib}$, 
the $2N_{\rm sib} \times 2N_{\rm sib}$  correlation matrix for $x_i$ is:
\begin{equation}
\label{eq:corr-matrix-pair}
R= \left(
\begin{array}{ccccc}
1 & r & 0 & 0 &     . \\
r & 1 & 0 & 0 &   . \\
0 & 0 & 1 & r &   . \\
0 & 0 & r & 1 &   . \\
.& .&. & . & . \\
\end{array}
\right)
\end{equation}
Each 2-by-2 sub-matrix in Eq.(\ref{eq:corr-matrix-pair}) represents 
a sibpair with off-diagonal element $r$ being the correlation coefficient 
$Cor(x_i, x_{i+1})$ between two sibs $i$ and $i+1$.  The variance of 
the extensive variable $X=\sum_i x_i$ is then equal to the weighted sum:
$$
Var_X  = \sum_{ij}  \sigma_i \sigma_j R_{ij}
$$
where $\sigma_i$ and $\sigma_j$ is the standard deviation of $x$ for person $i$ and $j$,
and the variance of the intensive quantity $x= \sum_i x_i/(2N_{\rm sib})$ is
$$
Var_x= \sum_{ij}  \sigma_i \sigma_j R_{ij}/(2N_{\rm sib})^2
$$
Since here we are dealing with sibpairs of the same affection status,
$\sigma_i=\sigma_j=\sigma$, which simplifies the variance for the 
correlation matrix in Eq.(\ref{eq:corr-matrix-pair}):
$$
Var_X = N_{\rm sib}  \cdot \sigma^2 \cdot 2(1+r) 
\hspace{0.2in}
and
\hspace{0.2in}
Var_x = \frac{ \sigma^2 (1+r)}{2 N_{\rm sib} }
$$
The equivalent number independent samples that lead to the same variance for $x$
can be derived by equating  $ \sigma^2 \cdot 2(1+r)/(2 N_{\rm sib}) = \sigma^2/N_E$, 
or, the ESS for sibpairs is:
\begin{equation}
\label{eq:ess-sib}
N_E= \frac{ 2N_{\rm sib}}{1+r}.
\end{equation}
The effective sample size reduction $\alpha$ is defined as the ratio between the ESS 
and the apparent sample size, and for sibpairs, it is equal to: 
\begin{equation}
\label{eq:alpha-sib}
\alpha \equiv \frac{N_E}{2N_{\rm sib}}= \frac{1}{1+r}.
\end{equation}

\vspace{0.1in}
\subsection*{Effective sample size for larger sibships} 

For $N_{\rm tri}$ pedigrees each with three siblings, the 
$3N_{\rm tri} \times 3N_{\rm tri}$ correlation matrix can be written as:
\begin{equation}
\label{eq:corr-matrix-sib3}
R= \left(
\begin{array}{ccccccc}
1 & r & r & 0 & 0 & 0 &  . \\
r & 1 & r & 0 & 0 & 0 &   . \\
r & r & 1 & 0 & 0 & 0 & . \\
0 & 0 & 0 & 1 & r & r &   . \\
0 & 0 & 0 & r & 1 & r &   . \\
0 & 0 & 0 & r & r & 1 &   . \\
. & . &.  & . & . &.  & . \\
\end{array}
\right)
\end{equation}
and the variance of $x$, ESS, and sample size reduction are: 
\begin{equation}
\label{eq:ess-sib3}
Var_x =  \frac{\sigma^2 (1+2r)}{3 N_{\rm tri}},
\hspace{0.15in}
N_E = \frac{ 3N_{\rm tri}}{1+2r }
\hspace{0.15in}
and
\hspace{0.15in}
\alpha= \frac{1}{1+2r}.
\end{equation}
More generally, for sibship of $k$ sibs, the sample size reduction  is
\begin{equation}
\label{eq:alpha-ksib}
\alpha= \frac{1}{ 1 + (k-1)r}.
\end{equation}

\vspace{0.1in}
\subsection*{Effective sample size for a mixture of relatives} 

For pedigrees with a specific mixture of relatives, for example, 
two sibs and one uncle, the correlation matrix consists of
identical sub-blocks:
\begin{equation}
R= \left(
\begin{array}{ccccccc}
1 & r_1 & r_2 & 0 & 0 & 0 & . \\
r_1 & 1 & r_2 & 0 & 0 & 0 & . \\
r_2 & r_2 & 1 & 0 & 0 & 0 & . \\
0 & 0 & 0 & 1 & r_1 & r_2 & . \\
0 & 0 & 0 & r_1 & 1 & r_2 & . \\
0 & 0 & 0 & r_2 & r_2 & 1 & . \\
. & . & . & .   &.    & . & . \\
\end{array}
\right) \nonumber
\end{equation}
where $r_1$ is the correlation coefficient between two sibs, and 
$r_2$ is that between a sib and the uncle. It can be shown that
the sample size reduction is
\begin{equation}
\label{eq:ess-mixture}
\alpha = \frac{3}{3+ 2r_1+4r_2}= \frac{1}{1+ 2 \overline{r} }
\end{equation}
where the averaged correlation $\overline{r}= (1/3) r_1 + (2/3) r_2$
is defined in such a way that we can assume all relatives were
similar and any two relatives have a correlation coefficient of $\overline{r}$.
The similar derivation can be generalized to any combination of relatives.

\vspace{0.1in}
\subsection*{Correlation coefficient of two relatives' allele counts} 

The correlation coefficient between allele count $x$ ($x$=2,1,0 for marker genotype 
$AA, AB, BB$, with probability of $p^2, 2pq, q^2$) of two sibs is:
$$
r \equiv \frac{ Cov[x_{\rm sib1}, x_{\rm sib2}] }{\sqrt{Var[x_{\rm sib1}]}
\sqrt{ Var[x_{\rm sib2}]} }
= \frac{ E[x_{\rm sib1}, x_{\rm sib2}]- E[x]^2 }{ Var[x]}.
$$
The mean and variance of the number of alleles is
$E[x]=2p$, $Var[x]=2pq$, and the joint probability $E[x_{\rm sib1}, x_{\rm sib2}]$
can be calculated by the Li-Sacks conditional probability given
the identity-by-descent  (IBD) status
\citep{lisacks,li98,li00,dai}.
The three Li-Sacks matrices (the so called ITO matrices) are the probability
of the second relative to have one of the genotypes given the genotype
of the first relative, and given the IBD status between the two relatives:
\begin{eqnarray}
\mbox{IBD=2} & \mbox{IBD=1} & \mbox{IBD=0}
        \nonumber \\
I=\left( \begin{array}{ccc}
1 & 0 & 0 \\ 0 & 1 & 0 \\ 0 & 0 & 1
\end{array} \right),
&
T=\left( \begin{array}{ccc}
p & q & 0 \\ p/2 & 1/2 & q/2 \\ 0 & p & q
\end{array} \right),
&
O=\left( \begin{array}{ccc}
p^2 & 2pq & q^2 \\ p^2 & 2pq & q^2 \\ p^2 & 2pq & q^2
\end{array} \right). \nonumber
\end{eqnarray}
By using the ITO matrices, we have ($\pi_k$ is the probability
of $k$ copies of IBD alleles between two relatives):
\begin{eqnarray}
E[x_{\rm rel1}, x_{\rm rel2}] &=& \sum_{i,j=0}^2 i \cdot j \cdot 
P(x_{\rm rel1}=i, x_{\rm rel2}=j)
= \sum_{i,j=0}^2 ij \sum_{k=0}^2 P(x_{\rm rel2}=j|i, k) \pi_k P(x_{\rm rel1}=i)
  \nonumber \\
&=& \pi_0 4p^2 + \pi_1 (4p^2+pq)+ \pi_2 (4p^2+2pq)
= 4p^2 + pq (\pi_1 + 2\pi_2). \nonumber
\end{eqnarray}
Inserting it back to the correlation coefficient formula, we have:
\begin{equation}
r= \frac{4p^2 + pq (\pi_1 + 2\pi_2) - 4p^2}{2pq} =\frac{\pi_1}{2}+\pi_2
 \nonumber
\end{equation}
The probability that a randomly selected allele from one relative is IBD with
a randomly selected allele from another relative, called kinship coefficient $\Phi$,
is equal to $\Phi= \pi_2 (1/2)+ \pi_1 (1/4) $ 
\citep{malecot,lange}. 
 The correlation coefficient $r$ is twice the value 
of kinship coefficient: $r= 2\Phi$. The same relationship was derived 
more tediously in, e.g., \citep{broman}  without using the ITO matrices.

\vspace{0.1in}
\subsection*{Correlation coefficient of two relatives' genotype indicator variable} 

Genotype indicator variable $x$ is 1 for a particular genotype of interest,
and 0 for other genotypes. For example, $x$=1,0,0 for $AA, AB, BB$ is
the indicator variable for the homozygous genotype $AA$. Using the
same ITO matrices, the joint probability for $AA$-indicator variable $x$
between two relatives is
\begin{eqnarray}
E[x_{\rm rel1}, x_{\rm rel2}] &=& P(x_{\rm rel1}=1, x_{\rm rel2}=1)
= \sum_{k=0}^2 P(x_{\rm rel2}=1|1, k) \pi_k P(x_{\rm rel1}=1) \nonumber \\
&=& p^2( \pi_2  + \pi_1 p  + \pi_0 p^2) 
 \nonumber
\end{eqnarray}
and correlation coefficient is:
\begin{equation}
\label{eq:cc-aa-indi}
r_{\rm AA-indicator} = \frac{ p^2( \pi_2  + \pi_1 p  + \pi_0 p^2) - (p^2)^2}{ p^2(1-p^2)}= 
\frac{ \pi_2  + \pi_1 p  + (\pi_0-1)  p^2}{1-p^2} 
\end{equation}
Similarly, for $AB$ and $BB$ indicator variable,
\begin{eqnarray}
\label{eq:cc-ab-indi}
r_{\rm AB-indicator} &=& \frac{ \pi_2  + \pi_1/2  + (\pi_0-1)  2pq}{1-2pq}
 \nonumber \\
r_{\rm BB-indicator} &=& \frac{ \pi_2  + \pi_1 q  + (\pi_0-1)  q^2}{1-q^2}
\end{eqnarray}

\vspace{0.1in}
\subsection*{Correcting $X^2$ test statistic and 95\% confidence interval
of odds-ratio by the effective sample size}

Single-marker case-control association analysis can be carried out
with chi-square test, odds-ratio (OR), and confidence interval of OR.
Typically, the control samples are randomly collected from a normal population
with no need for correcting correlated samples, whereas case samples
might be collected during the linkage analysis stage, thus are correlated.
For allele-based analysis \citep{sasieni}, denote the allele counts in
case group  as $N_{\rm A,case}, N_{\rm B,case}$ and those in control group
as $N_{\rm A,con}, N_{\rm B,con}$, the Pearson's chi-square test statistic can
be recalculated by replacing $N_{\rm A,case}, N_{\rm B,case}$ with
$\alpha N_{\rm A,case}$ and $\alpha N_{\rm B,case}$:
\begin{equation}
\label{eq:x2-new}
X^2_e = \frac{ \alpha (N_{\rm A,case}N_{\rm B,con}- N_{\rm B,case}N_{\rm A,con})^2
(\alpha N_{\rm A,case} + \alpha N_{\rm B,case}+N_{\rm A,con}+N_{\rm B,con})}
{(N_{\rm A,case}+N_{\rm B,case})(N_{\rm A,con}+N_{\rm B,con})
 (\alpha N_{\rm A,case}+N_{\rm A,con}) (\alpha N_{\rm B,case}+N_{\rm B,con})}.
\end{equation}
The modified test statistic $X^2_e$ can then be used to determined the $p$-value.

For OR $\hat{\theta} = N_{\rm A,case} N_{\rm B,con}/(N_{\rm A,con}N_{\rm B, case})$,
the uncorrected 95\% confidence interval (CI) is estimated by the Woolf's formula
\citep{woolf}:
$ [l, u]= [ e^{ \log \hat{\theta} - 1.96 \hat{ \sigma} ( \log \hat{\theta}  ) },
e^{ \log \hat{\theta} + 1.96 \hat{ \sigma} ( \log \hat{\theta}  ) }]$,
with $ \hat{ \sigma} ( \log \hat{\theta}  ) =
( 1/N_{\rm A, case} + 1/N_{\rm B,case} +
1/N_{\rm A, con} + 1/N_{\rm B,con} )^{0.5}. $
This can be corrected in a similar way by replacing 
$N_{\rm A,case}, N_{\rm B,case}$ with $\alpha N_{\rm A,case}$ and 
$\alpha N_{\rm B,case}$:
\begin{equation}
\label{eq:or-new}
\hat{ \sigma}_e ( \log \hat{\theta}  ) =
\left(
\frac{1}{\alpha N_{\rm A, case}} + \frac{1}{\alpha N_{\rm B,case}} +
\frac{1}{N_{\rm A, con}} + \frac{1}{ N_{\rm B,con}}
\right)^{1/2}.
\end{equation}
It can be shown that $\alpha < X^2_e/X^2 < 1$ and $\hat{ \sigma}_e/\hat{ \sigma} > 1$,
when $\alpha < 1$. In other words,  when the effective sample size is 
smaller than the apparent sample size, the test statistic is smaller 
(leading to larger $p$-values), and the 95\% CI of OR is wider.

\section*{Results}

\subsection*{Diminishing return in adding more relatives from the same pedigree in
an association study} 

The kinship coefficients and sample size reduction with respect to allele 
frequency estimation of common relative pairs are listed in Table \ref{table:rel}, and those for 
sibships with 1,2, $\dots$ siblings are listed in Table \ref{table:sib}.
For more complicated relationships or pedigrees with loop, one can consult
\citep{maruyama,lange}.  Several rules-of-thumb can be stated:
two siblings contribute 1.333 samples, uncle-nephew pair contributes 1.6
samples, three siblings are equivalent to 1.5 samples, etc. If the relationship
between two pedigree members is distant, the correlation is close to zero
and they can be treated as two independent samples (e.g., second cousins
contribute 1.94 samples). 
For larger sibship, there is a diminishing return in adding
extra sibs: adding the second, the third, the fourth, and the fifth sibs
only adds 0.333, 0.167, 0.1, 0.067 samples. Even in the limit of infinite
number of sibs, the effective sample size can not be larger than 2,  as 
the extra sibs merely resample the finite pool of four parental alleles. 

These results show that while one should include as many samples as possible,
whether correlated or not, in an association study, it does not seem necessary
to include too many relatives from the same pedigree. While distant
relatives are essentially independent samples, for close relatives such as 
siblings, two persons are perhaps a good compromise between the desire 
to add more samples and the diminishing return due to correlations.

When a mixture of relatives from the same pedigree is included, one
can use the averaged correlation coefficient discussed in the
Method section. For example, with two siblings and one aunt/uncle, the
averaged correlation coefficient $\overline{r}= (1/3)0.5 + (2/3)0.25 =1/3$.
The ESS for the two-sib-one-uncle is 1.8, larger than the value of 1.6 
for three siblings.

\begin{table}[ht]
\begin{center}
\begin{tabular}{|c|ccc|c|c|c|c|}
\hline
pair relationship & $\pi_2$ & $\pi_1$ & $\pi_0$ & $\Phi$ & $r$ & $\alpha$ & $N_E$\\
\hline
parent-child& 0 & 1 & 0  & 1/4 & 1/2 &  2/3 & 4/3 $\approx$ 1.333\\
sibs & 1/4 & 1/2 & 1/4 & 1/4 & 1/2 & 2/3  & 4/3 $\approx$ 1.333\\
half-sibs  & 0 & 1/2 & 1/2  & 1/8 & 1/4 & 4/5  & 8/5 = 1.6 \\
uncle/aunt-nephew/niece& 0 & 1/2 & 1/2 & 1/8 & 1/4 & 4/5 & 8/5 =1.6 \\
first cousins & 0 & 1/4 & 3/4 & 1/16 & 1/8 & 8/9  & 16/9 $\approx$ 1.778\\
second cousins & 0 & 1/16 & 15/16 & 1/64 & 1/32 & 32/33 & 64/33 $\approx$ 1.939 \\
\hline
\end{tabular}
\caption{\label{table:rel}
For several common relative pairs, these quantities are listed:
$\pi_2, \pi_1, \pi_0$: probabilities of 2,1,0 copies of allele that
are identity-by-descent (IBD); $\Phi$: kinship coefficient;
$r$: correlation coefficient between the number of allele $A$ (or $B$)
counts; $\alpha$: sample size reduction; $N_E$: effective number
of samples in the relative pair.
}
\end{center}
\end{table}

\begin{table}[ht]
\begin{center}
\begin{tabular}{|c|c|c|}
\hline
size of sibship &  $\alpha$ & $N_E$ \\
\hline
2 &  2/3 & 4/3 $\approx$ 1.333 \\
3 & 1/2 & 3/2=1.5 \\
4 & 2/5 & 8/3=1.6 \\
5 & 1/3 & 5/3 $\approx$ 1.667 \\
$k$ & 2/($k$+1) & $2k/(k+1) \approx 2(1-1/k) $ \\
\hline
\end{tabular}
\caption{\label{table:sib}
The sample size reduction $\alpha$ and effective sample size $N_E$
of sibships with 2, 3, 4, 5, and $k$ sibs.
}
\end{center}
\end{table}

\vspace{0.1in}
\subsection*{Improving $p$-value by using all samples} 

\begin{table}[ht]
\begin{center}
\begin{tabular}{cccccc}
\hline
 & TT & TC & CC  & N & $N_{\rm allele}$=2N\\
\hline
case (86 singletons and 377 sibpairs) & 21 & 241 & 578  & 840 & 1680 \\
case (86 singletons and 377 sibs) & 10 & 126 & 327 & 463 & 926 \\
control & 9 & 143 & 774 & 926 & 1852 \\
\hline
\end{tabular}
\caption{\label{table:ptpn}
Genotype counts of a SNP in {\sl PTPN22} in human chromosome 1 in case
(rheumatoid arthritis) and control group. The first line summaries
the genotype counts of all case samples, including 86 singletons
(uncorrelated samples) and 377 sibpairs. The second line is a subset
of the case group with one affected sib per pedigree (sibpair) randomly
chosen. The third line is for the control group.
}
\end{center}
\end{table}
 
For the {\sl PTPN22} data in Table \ref{table:ptpn}, if one affected sib
per sibpair is selected for association as in \citep{begovich},
$X^2=31.42$ leads to $p$-value of 2.1 $\times 10^{-8}$ (with Fisher's 
exact test, the $p$-value is 5.6 $\times 10^{-8}$), and 95\%CI of OR is 
(1.55-2.50). We know this is an underuse of the samples as the second sibs 
in sibpairs were discarded. Using all sibs in sibpairs without correction 
leads to the incorrect result of $X^2=$53.26, $p$-value of 2.9 $\times 10^{-13}$,
and 95\% CI of OR of (1.73-2.62).  The overall ratio of effective sample 
size and the apparent sample size is: 
$\alpha=$ (86 + 377 $\times$ 2 $\times$ 2/3)/(86 + 377 $\times$ 2) $\approx$ 0.70.
Using the Eq.(\ref{eq:x2-new}) and Eq.(\ref{eq:or-new}), the modified 
$X^2=45.73$ leads to $p$-value of 1.36 $\times 10^{-11}$, and modified 
95\% CI of OR (1.70-2.66). Compared to the one-sib-per-pair dataset, 
even though the conclusion on statistical significance is unchanged, 
the $p$-value is 1500 times smaller. 

The ratio of two chi-squares, one for all samples with ESS correction
and another without, is calculated to be $X^2_e/X^2= 45.73/53.26 =0.86$.
This ratio can also be approximately estimated from ESS. 
Since $X^2$ and $X^2_e$ can be written in the form: 
$X^2= (\hat{p}_{\rm A, case}- \hat{p}_{\rm A,control})^2/[(1/N_{\rm case} + 
1/N_{\rm control}) \cdot \overline{p} \cdot \overline{q}$],
$X^2_e \approx (\hat{p}_{\rm A, case}- \hat{p}_{\rm A,control})^2/
[(1/(\alpha N_{\rm case}) + 1/N_{\rm control})
\cdot \overline{p} \cdot \overline{q}]$ (in an approximation, the pooled 
allele frequency estimation for A and B is not greatly affected by 
the change of sample size),
$X^2_e/X^2 \approx (1/(0.7 \times 1680) + 1/1852)/(1/1680+1/1852)= 0.82$.

\begin{table}[ht]
\begin{center}
\begin{tabular}{cccccc}
\hline
 & CC & TC & TT & N & $N_{\rm allele}$=2N\\
\hline
case (all) & 169 & 624 & 535 & 1328 & 2656 \\
case (1 sample per ped) & 87 & 308 & 258 & 653 & 1306 \\
control & 247 & 603 & 494  & 1344 & 2688 \\
\hline
\end{tabular}
\caption{\label{table:ifih1}
Genotype counts of a SNP in {\sl IFIH1} gene in human chromosome 2 in case
(rheumatoid arthritis) and control group. Those of all case samples,
of independent case samples, and of control samples, are listed in
lines 1,2, and 3.
}
\end{center}
\end{table}

For the {\sl IFIH1} gene in Table \ref{table:ifih1}, we applied the effective 
sample size method both globally or pedigree-type-specifically. 
Using Eqs.(\ref{eq:ess-sib},\ref{eq:ess-sib3},\ref{eq:alpha-ksib},\ref{eq:ess-mixture}),
and by a conservative use of relatives in assuming all relatives to
be sibships, we have the averaged effective number of allele counts:
$2(67+ 512 \cdot 2/(1+0.5)+ 64 \cdot 3/(1+2 \times 0.5)+ 8 \cdot 4/(1+3 \times 0.5)+  5/(1+4 \times 0.5)
+ 8/(1+ 7 \times 0.5) \approx 1724$, or,
the average sample reduction of $\alpha= 861.9111/1328 \approx 0.649$.
The ESS-based method leads to a $p$-value of 0.0023 in chi-square test,
improved upon the $p$-value of 0.0179 when only one case per family 
is used (second line in Table \ref{table:ifih1}). At the significance level
of 0.01, adding correlated samples in this dataset makes an 
insignificant result significant.

The SNP minor allele frequency (MAF) for the control population in the
{\sl IFIH1} gene was reported to be 40.4\% in the initial round,
and 38.7\% in the follow-up round \citep{smyth}; that
in the type 1 diabetes population was 35.3\% in the first
round, and 34.0\% in the second round, each with thousands of
samples. The control MAF in Table \ref{table:ifih1}
is 40.8\%, consistent with the value in \citep{smyth}.
However, the MAF for the rheumatoid arthritis samples in
Table \ref{table:ifih1} is 36.2\%, larger than the MAF for the
type 1 diabetes samples. The weaker association signal in
rheumatoid arthritis study as compared to diabetes study
implies a larger sample size requirement for its detection,
and as a result, ESS method proves to be important in
incorporating extra sibling samples to increase the sample size.

One can also apply the ESS method to each pedigree-type specifically. We
count the T and C alleles in pedigrees with only affected sibpairs,
then reduce the count by the factor $1/(1+0.5)=2/3$. Similarly, the
allele counts in pedigrees with three affected sibs are reduced by
the factor of $1/(1+2 \times 0.5)=1/2$, etc. The pedigree-type-specific
allele count reduction leads to $p$-value of 0.00467. We can partially
explain why this $p$-value is not as good as the one derived by the global
sample size reduction:  the association signal is largely due to
an enrichment of the major allele T in the case group; however,
the largest pedigree with 8 affected members with 13 counts 
of the T allele leads to an effective contribution in the ``local 
method" of 3.3 counts, as versus the 9.7 counts in the ``global method".

\vspace{0.1in}
\subsection*{A single effective sample size does not capture all
variance inflations in genotype frequency estimations, but it
provides a good approximation} 

With the correlation coefficient for genotype indicator variable
in Eq.(\ref{eq:cc-aa-indi},\ref{eq:cc-ab-indi}), we can derive the
sample size reduction $\alpha$ and variance inflation $1/\alpha$
for genotype frequencies obtained from relative pairs, sibships, 
and cluster of relatives: $\alpha_G=1/(1+r_G)$, $1/( 1+(k-1) r_G)$,
and $1/(1+ (k-1) \overline{r}_G)$ respectively, where $r_G$ (G=(AA,AB, BB)
is the genotype-specific correlation coefficient.  Compared to the 
variance inflation for allele frequency estimation, the number of ESSs
for genotype frequencies is 3 instead of 1, as $r_{AA}, r_{AB}, r_{BB}$ 
are not equal to each other. 
Furthermore, these correlation coefficients depend on $p,q$.

\begin{figure}[th]
  \begin{turn}{-90}
  \epsfig{file=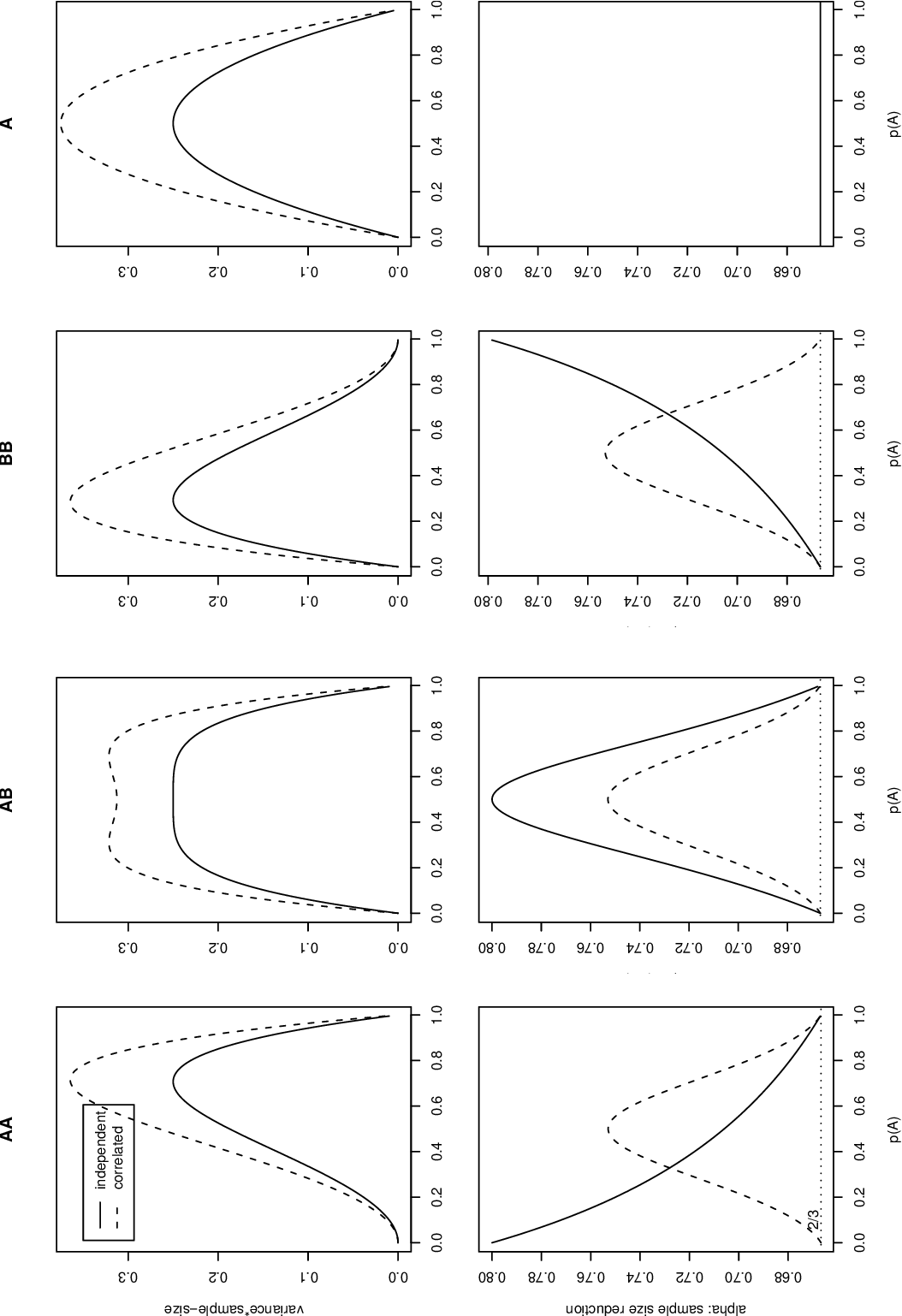, width=11cm}
  \end{turn}
\caption{
\label{fig:three-alpha}
(Upper row) expected variance of genotypes AA, AB, BB and allele A (multiplied
by the sample size) as a function of the allele frequency $p(A)$. The solid 
line indicates the result from independent samples, and dashed line from
sibpairs.
(Lower row) effective genotype count reduction $\alpha_1, \alpha_2, \alpha_3$ 
for sibpair data as a function of $p(A)$ (Eq.(\ref{eq:alpha-geno})). 
For allele count, the sample size reduction is a constant number of 2/3.
The grey line is the $\alpha_a(p)$, the weighted average of
$\alpha_1, \alpha_2, \alpha_3$. The $\alpha$=0.7096 line is the
average of $\alpha_a(p)$ over $p$'s.
}
\end{figure}

We illustrate these properties by the example of sibpairs. Using
Eq.(\ref{eq:cc-aa-indi},\ref{eq:cc-ab-indi},\ref{eq:alpha-sib}),
the genotype-specific sample size reductions are:
\begin{eqnarray}
\label{eq:alpha-geno}
\alpha_{\rm AA, sibpair} &=& \frac{1}{1+ (1+3p)/(4+4p)} \nonumber \\
\alpha_{\rm AB, sibpair} &=&  \frac{1}{1+ (1-3pq)/(2-4pq)} \nonumber \\
\alpha_{\rm BB, sibpair} &=&  \frac{1}{1+ (1+3q)/(4+4q)}.
\end{eqnarray}
Figure \ref{fig:three-alpha} shows $\alpha_{\rm G, sibpair}$'s of the three
genotypes  as a function of $p$; also shown are the genotype
frequency variance (multiplied by the sample size).
Variances of genotype frequencies calculated from independent samples
are shown in solid lines as a comparison. It can be seen from 
Figure \ref{fig:three-alpha} that the variance inflation of allele 
frequency is distinct from those of genotype frequencies in that 
its $\alpha$ is a constant value 2/3 independent of $p$. It 
illustrates that one should not expect a single parameter to correct 
the variance inflation in correlated samples for all circumstances.

The genotype-specific sample size reductions in Eq.(\ref{eq:alpha-geno})
can be applied in the following way: (1) the allele frequency $p$
is estimated from the data; (2) three $\alpha_G$s are calculated
by Eq.(\ref{eq:alpha-geno});
(3) each genotype count is discounted by the genotype-specific
$\alpha_G$, then these modified genotype counts can be used
for further genotype-based association.  Notice that $\alpha_G$'s 
in Eq.(\ref{eq:alpha-geno}) are confined to the range (2/3, 4/5). 
One can also obtain an averaged ESS by averaging over three genotypes:
$\alpha_{\rm avg, sibpair}(p)= p^2 \alpha_1 +2pq \alpha_2 + q^2 \alpha_3$,
and $\alpha_{\rm avg, sibpair}$ ($p$ is estimated from the data first)
can be used to discount all three genotype counts by the same
factor. In yet another approach, $\alpha_{\rm avg, sibpair}$
can be averaged over $p$: $\overline{\alpha} \equiv \int_0^1 \alpha_a(p) dp$. 
This leads to $\overline{\alpha}=0.7096$. One can use $\overline{\alpha}$
to discount all three genotype counts without the need to estimate
$p$ first. Note that this sample size reduction
is less severe than that to account for variance inflation in
allele frequency estimation, $\alpha=0.6667$.

\begin{table}[ht]
\begin{center}
\begin{tabular}{ccccccc}
\hline
 & & \multicolumn{2}{c}{$\alpha=0.01$} & &\multicolumn{2}{c}{$\alpha=0.05$} \\
\cline{3-4} \cline{6-7}
$p$ & model & using $X^2$ & using $X^2_e$ & & using $X^2$ & using $X^2_e$ \\
\hline
0.1 & R & 0.022 & 0.011   & & 0.082 & 0.054  \\
    & A & 0.022 & 0.011   & & 0.083 & 0.051  \\
    & D & 0.019 & 0.009   & & 0.078 & 0.051  \\
0.3 & R & 0.022 & 0.011   & & 0.073 & 0.048  \\
    & A & 0.020 & 0.009   & & 0.078 & 0.049  \\
    & D & 0.022 & 0.010   & & 0.082 & 0.052  \\
0.5 & R & 0.022 & 0.011   & & 0.082 & 0.053  \\
    & A & 0.020 & 0.009   & & 0.078 & 0.050  \\
    & D & 0.022 & 0.010   & & 0.082 & 0.053  \\
\hline
\end{tabular}
\caption{\label{table:t1}
Empirical type I errors for the allele-based association test
at the nominal significance level of 0.01 and 0.05, either by 
using the naive (uncorrected) $X^2$ or the $X^2_e$ modified by 
the effective sample size. The allele frequencies in case and 
control group are the same, even though different simulation runs 
are labeled as recessive (R), additive (A) and dominant (D).
}
\end{center}
\end{table}

\begin{figure}[th]
  \begin{turn}{-90}
  \epsfig{file=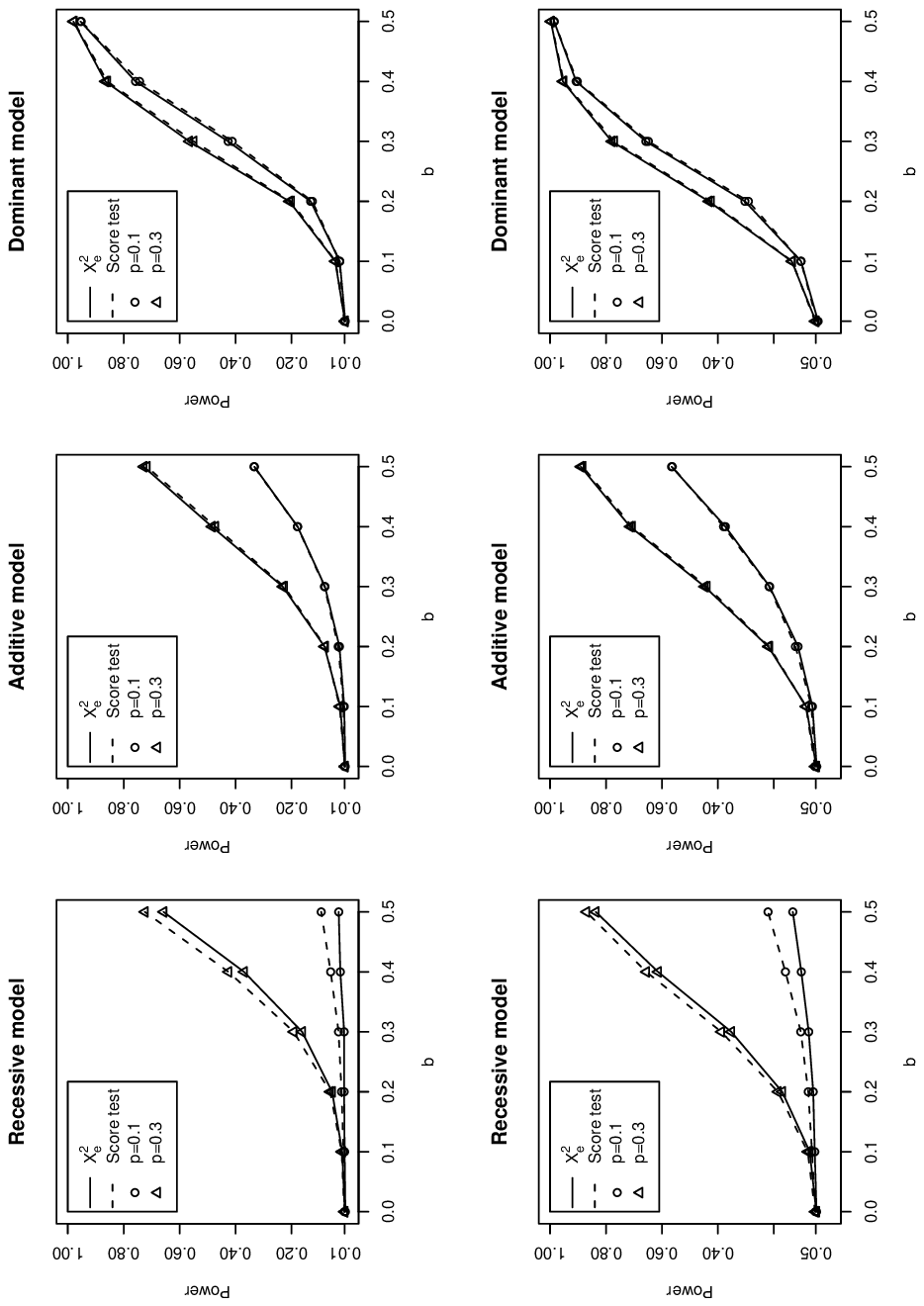, width=11cm} 
  \end{turn}
\caption{
\label{fig:power-score}
Empirical power curve for the genotype-based test of three different
models (recessive, additive, dominant) at the nominal significance
level of 0.01 (upper row) and 0.05 (lower row). The $x$-axis is the
log-odds ratio parameter $b$ in the disease model Eq.(\ref{eq:simu-model}). 
Two power curves are shown: using effective sample size corrected 
$X^2_e$  (solid line), and by the score test (dashed line).
}
\end{figure}

\vspace{0.1in}
\subsection*{Effective sample size method performs well
in simulation and in comparing the score test}

Using the simulated data described in the Methods and Material section,
we have checked the validity of the effective sample size method. We 
first compare the test errors in using the uncorrected chi-square 
test statistic $X^2$ and the ESS-corrected $X^2_e$, for the allele-based
test.   Table \ref{table:t1} shows the type I error under the null distribution in chi-square
test using the naive $X^2$ and ESS-corrected $X^2_e$. Note
that for the null distribution, different disease model has no effect 
on the allele/genotype frequencies, and we simply consider the 
R/A/D models in Table \ref{table:t1} as three independent runs.
It is clear that $X^2_e$ leads to the more correct type I errors,
practically identical to the nominal significance, whereas the
naive $X^2$ clearly leads to larger type I errors.

The locally most powerful test among all tests with the correct type I errors
is the score test \citep{cox}  which sets a standard other
tests can be compared to. For allele-based analysis, ESS-corrected $X^2_e$
is identical to the score test, sharing the same power. 
For genotype-based test (i.e. chi-square test on 2-by-3 genotype count
table), chi-square test using ESS-corrected $X^2_e$ is not identical
to the score test. Here we adopt the simplest ESS correction
for genotype data: multiplying the genotype counts by a constant
reduction value $\overline{\alpha}=0.7096$.  The power curve in 
Figure \ref{fig:power-score} shows that the difference between the
ESS-corrected $X^2_e$ test and score test is negligible for dominant 
or additive disease models. The difference for recessive models 
is non-zero, but nevertheless small.

\section*{Discussion}

{\bf Cheverud's formula for the number of independent variables }
Based on the idea that the overall amount of correlation among
several variables can be measured by the variance of the
eigenvalues derived from their correlation matrix, Cheverud
proposed a formula to calculate the effective number of variables 
\citep{cheverud}:
\begin{equation}
\label{eq:cheverud}
N_E = N \left( 1- (N-1) \frac{Var[\lambda]}{N^2} \right)
\end{equation}
where $N$ is the number of variables, and $\lambda=(\lambda_1, \lambda_2, \dots
\lambda_N)$ are the eigenvalues of the $N \times N$ correlation matrix 
for these $N$ variables. Eq.(\ref{eq:cheverud}) has been applied to 
QTL mapping in the inbreeding system and to human association analysis to determine
the number of independent markers in a linkage disequilibrium block
\citep{cheverud,dale}.  Although this formula has not been used
to determine the number of independent samples, it can be interesting 
to compare Eq.(\ref{eq:cheverud}) with the ESS formula derived in this paper.

We consider the large sibship situation where the correlation matrix
is characterized by Eq.(\ref{eq:corr-matrix-sib3}). It can be shown
that each submatrix (the $3 \times 3 $ block in Eq.(\ref{eq:corr-matrix-sib3}))
contributes an eigenvalue equal to $1+(3-1)r=1+2r$, and two eigenvalues 
equal to $1-r$.
The variance of the eigenvalues for Eq.(\ref{eq:corr-matrix-sib3}) is
then equal to $2r^2 (N/(N-1))$. Inserting it to Eq.(\ref{eq:cheverud}), we have
the Cheverud's effective number of variables: $N-2r^2$. Compared to
the ESS of $N/(1+2r)$ determined by Eq.(\ref{eq:ess-sib3}), Cheverud's
formula leads to a larger effective number of degrees of freedom,
and less reduction, in particular in the large $N$ limit.

We believe that our effective sample size formula makes better sense:
in the three-sib sibship case, because each sibship is independent
of another, the number of independent samples is at least $N/3$. 
Note that the ESS formula involves an operation of rescaling the 
original sample size N, instead of subtracting a correction term. 
In order for Cheverud's formula to have a similar effect, the 
variance of eigenvalues has to increase with the sample size $N$. This can 
be true only if there is a collective correlation for all variables,
or if there is haplotype block-block correlation. If the variables 
(samples) can be split into independent blocks, the effective 
degrees of freedom (sample size) should always be a rescaled
version of the original one.  Interestingly, for a model discussed 
in \citep{salyakina} where the correlation coefficient within a 
block is 1 and those between blocks are small non-negative values, 
the effective number of variables is indeed a rescaled value of the 
original number of variables.

\vspace{0.1in}
{\bf The variance inflation factor in the genomic control method}
The genomic control method in association studies was proposed in
\citep{devlin99,bacanu,devlin01} to correct 
population stratification and ``cryptic relatedness" between samples.
Despite quantitative differences in the mechanism for correlation, population 
stratification and family clusters could have similar consequences,
and this similarity is exploited in a unified framework for
association studies of quantitative traits \citep{yu}.
In the genomic control method, neutral markers are used to estimate
the variance inflation factor $\lambda$, and $\lambda$ is used to divide
the chi-square statistic: $X^2_{\rm gc}= X^2/\lambda$ for a modified
test statistic. This can be compared to our formula for an ESS-corrected 
chi-square test statistic in Eq.(\ref{eq:x2-new}), $X^2_e \approx \alpha X^2$
(if the allele counts $N_{\rm A,con}, N_{\rm B,con}$ in control group are not too small).
In this approximation, $X^2_e \approx X^2_{\rm gc}$  if $\alpha=1/\lambda$.

Whether genomic control can correctly capture the
population substructures is still under debate \citep{devlin04},
with reports of either under- or over-correcting the correlation depending 
on the number of markers used \citep{marchini,kohler},
and its performance perhaps also depends on whether the markers used to 
estimate $\lambda$ are ancestral-informative or not. For whole genome
association studies with a large number of markers, it is recommended
to use a Bayesian version of the genomic control \citep{devlin04}. 
In our situation, we are correcting the known relatedness 
between samples, and there is no issue of under- and over-correcting 
the test statistic.

One key debate on genomic control is whether $X^2_{\rm gc}$ follows
a central or non-central chi-square distribution \citep{gorroochurn}.
For a truly admixed population with a positive Wright's $F_{ST}$
value, the variance of the allele frequency is $Var_p= p(1-p) (F_{ST} - F_{ST}/N+1/(2N) )$
(the inbreeding coefficient is assumed to be zero) \citep{weir},
which is inflated by a factor $(F_{ST} - F_{ST}/N+1/(2N) )$.
This admixture-induced variance inflation cannot be accounted for by 
a simple sample size reduction, because even of the infinite sample 
limit the residual variance is still nonzero. At the infinite sample
size  limit, the variance inflation factor is equal to $F_{ST}$,
which is why $F_{ST}$ is also called the standardized measure of variance, 
or Wahlund's variance \citep{cavalli}.
The only way to reconcile the variance inflation and sample size
reduction here is to set $\alpha = 1/(1 + 2(N-1) F_{ST})$, i.e.,
the sample size reduction itself depends on sample size.
All these issues in correcting admixed subpopulations are
not problematic for our relative samples because we assume
the allele frequency does not change from pedigree to pedigree.

\vspace{0.1in}
{\bf Comparison to the generalized estimation equation approach} 
The method of generalized estimation equation (GEE), similar to ESS
method, has a goal of utilizing correlated samples in an analysis
\citep{liang,hanley}.  However, one major difference 
between GEE and ESS is that GEE relies on data to estimate the within-cluster
correlation among samples, whereas ESS calculates the correlation
by the information given. Typically in GEE, only a single correlation coefficient
$r$ is estimated for all clusters, which can be unreliable if clusters
of different natures are included in the data. For example, if the dataset
contains both sibpairs and cousin-pairs, the $r$ for samples within sibpairs
should be larger than that for cousin-pairs. Another difference is that
GEE corrects not only variance, but also mean as well, whereas ESS only
modifies variance. Similar to an argument made in \citep{devlin04},
we believe that sample correlation mainly affects the variance, and has
less effect on bias. 

We use the {\sl IFIH1} genotype data in Table \ref{table:ifih1} to illustrate differences between
GEE and ESS.  Using the {\sl corstr=``exchangeable"} option in the {\sl gee} 
subroutine in $R$ statistical package (VJ Carey, T Lumley, and B Ripley, 
``The {\sl gee} package", version 4.13-12, Feb 2007) , the 
averaged within-family correlation coefficient for the
allele count variable was estimated as $r=$0.4349. This $r$ value is slightly
smaller than that for sibpairs ($r=$0.5), but close, reflecting the fact that
this dataset is dominated by sibpairs.  In the Results section,
we have shown that using $r=0.5$, the sample size reduction for the dataset
in Table \ref{table:ifih1} is equal to 0.649.  If we use the within-group 
correlation coefficient $r=$0.4349  estimated by GEE, the sample size reduction 
is 0.678. The GEE and ESS  results are more or less the same, though GEE
does not seem to correct the correlation enough. A similar observation
that GEE tends to underestimate variance for smaller sample sizes
was made in \citep{tregouet}.

The estimation equation is essentially a procedure to determine
weights of samples. When samples are correlated, their weights
are lower than 1. It was shown in \cite{hanley} that the weight 
for sibpairs $w$ that minimizes the variance is exactly equal to 
the Eq.(\ref{eq:alpha-sib}) used in this paper. We expect that
in general, the weight of related samples determined by minimizing 
the variance will be equal to the sample size reduction 
$\alpha$ if the weight for independent individuals is set to 1.
 
In conclusion, among alternative approaches in handling correlated
samples in genetic association studies, such as likelihood-based
approach \citep{bourgain}, sample weighting \citep{browning},
and estimation equation \citep{tregouet}, effective
sample size is perhaps the most accessible method: easier to use,
and with no need to have new computer software. Since the reason
that correlated samples are often avoided in practice is not
because solutions do not exist, but because the existing methods
are relatively hard to use, we believe the ESS method discussed here
will help medical geneticists to routinely use pedigree data in
association studies.

\section*{Methods and Materials}

\subsection*{Data sets}

A missense SNP {\sl rs2476601} in the protein tyrosine phosphatase non-receptor 
type 22 gene on chromosome 1 ({\sl PTPN22}) was shown to 
be associated with the autoimmune disease rheumatoid arthritis 
\citep{begovich,annette}. The rheumatoid arthritis samples 
were collected by the North American Rheumatoid Arthritis Consortium 
(NARAC) for genetic linkage analysis \citep{damini01,damini03},
and all pedigrees contain two or more affected siblings. 
In the original report \citep{begovich},  one sib per affected sibpair 
is randomly selected from 377 affected sibpairs for the association 
analysis (plus 86 singletons).  This procedure cuts the number of case samples
almost by half. An association 
analysis of all affected sibs with a correction of the correlation between sibs
was not carried out. We reproduce this dataset in Table \ref{table:ptpn}
(corresponds to the ``replication study" in Table 1 of \citep{begovich}).

Another dataset used here is the genotype of a non-synonymous SNP 
in {\sl IFIH1} gene on chromosome 2, also collected by NARAC. 
{\sl IFIH1} gene has recently been shown to be associated with type 1
diabetes \citep{smyth}, but its association status with
rheumatoid arthritis is unknown.
This dataset consists of 1344 independent control samples and 1328 case 
samples distributed in 653 pedigrees -- including 67 singletons, 
512 affected relative-pairs (the majority are affected sibpairs), 
64 affected triples (most are sibship with 3 affected sibs), 8 affected quadruples, 
and two pedigrees with 5 and 8 affecteds.  The three genotype counts  of
this SNP are listed in Table \ref{table:ifih1}.

\vspace{0.1in}
\subsection*{Simulations}

Simple simulated datasets were created for checking the effective 
sample size method as applied to sibpair data. For each replicate,  genotypes of
500 ``case" samples consisting of 250 sibpairs and 500 ``control" samples
were simulated. The genotype in control group was sampled from the
genotype distribution of $P_{\rm control}(G)=$ ($p^2, 2pq, q^2$) for 
genotypes AA, AB, BB. Those in the case group is sampled by the model:
\begin{equation}
\label{eq:simu-model}
P_{\rm case}(G) \propto  \frac{e^{a+b \cdot f(G)}}{1+e^{a+b \cdot f(G)}} P_{\rm control}(G)
\end{equation}
where $f(G)$ represents the disease models, $a$ is the baseline
log-odds, and $b$ is the log-odds ratio.
The dominant model (D) is equivalent to $f(G)$=(1,1,0) for genotype AA, AB, BB;
recessive model (R) corresponds to $f(G)$=(1,0,0); and additive model (A)
corresponds to $f(G)=$ (1, 0.5, 0). For the null distribution to be used to check 
the type I error, $b=0$, i.e, genotype has no effect on the disease 
status, and  $P_{\rm case}=P_{\rm control}$.  For the alternative 
distribution to be used to check the power, $a$ is chosen at $-4$ 
and $b$ is chosen between 0 and 0.5.

\section*{Acknowledgments}

WL and PKG are supported by the National Institute of Health 
grant R01-AR44422, NO1-AR22263. YY is supported by the National
Natural Science Foundation of China Grant 10671189 and the 
Chinese Academy of Science Grant No. KJCX3-SYW-S02.  EFR, 
CBO and DLK was supported in part by the Intramural Research Program 
of the National Institute of Arthritis and Musculoskeletal and 
Skin Diseases of the National Institutes of Health. We would like 
to thank Chris Amos, Yongchao Ge, Jianxin Shi, Jan Freudenberg
for helpful discussion.

\end{document}